# Ground Experiment of Full-Duplex Multi-UAV System Enabled by Directional Antennas

Tao Yu, Kiyomichi Araki, Kei Sakaguchi
Tokyo Institute of Technology, Japan
{yutao, araki, sakaguchi}@mobile.ee.titech.ac.jp

*Abstract*—A high performance multi-UAV communication system, which bridges multiple UAVs and ground station, is one of the key enablers to realize a variety of UAV-based systems. To address the issues such as the low spectrum efficiency caused by the co-channel interference, we have proposed a spectrum-efficient full-duplex multi-UAV communication system with low hardware complexity. In this paper, on-ground experiments are conducted to confirm the feasibility and effectiveness of the key feature of the proposed system, i.e., co-channel interference cancellation among UAVs by directional antennas and UAV position control, instead of energy-consuming dedicated self-interference cancellers on UAVs in traditional full-duplex systems. Channel power of interference link between a pair of two UAVs reusing the same channel is measured, and the achievable channel capacity is also measured by a prototype system implemented by software-defined radio devices. The results of different antennas and different antenna heights are also compared. The experimental results agree well with the designs and confirm the feasibility and effectiveness of the proposed system. This ground experiment is a work in progress to provide preliminary results for the multi-UAV-based experiments in the air in the future.

## I. INTRODUCTION

In recent years, terrestrial and aerial robots have evolved to be more operable, functional, productive, and affordable than ever before thanks to the increasing miniaturization of electronic components and the developments of high-performance control algorithms. Capabilities such as autonomous mobility and unattended operation make robot systems suitable mobile platforms for varieties of unmanned applications. The merits are especially magnified for multi-robot systems, in which multiple robots can collaboratively execute complicated tasks, such as sensing, rescue, and exploration [1][2]. Multiple unmanned aerial vehicle (UAV) systems especially gain our research interest because they further extend robot mobility from 2D to 3D space.

High-performance communication is one of the key technologies to realizing multi-UAV systems and related applications because it enables real-time information exchange between multiple UAVs and ground stations (GSs) via wireless communication. However, there are still several issues that are hindering the development of high-performance multi-UAV communication systems, such as follow. 1) The lack of obstacles in the aerial propagation environment leads to severe co-channel interference (CCI) among UAVs, which results in low system performance and spectrum efficiency. 2) UAVs have limited battery-life and payload, so they are desired to be of low hardware complexity and low energy consumption, which puts a significant limitation on the functionality of functional modules on UAVs, including the communication system.

Currently, most multi-UAV communication systems are based on existing standardized techniques such as IEEE 802.11 families, cellular families, and even IEEE 802.15 families, or their derivatives [3]-[5]. All of them are out-band full-duplex (OBFD) systems, i.e., they cannot perform transmission and reception with the same radio resources, which results in low radio resource efficiencies, especially in systems with multiple UAVs. Therefore, the usages of in-band full-duplex (IBFD) have also been explored in UAV-based applications, such as full-duplex UAV relays [6]-[9], but they suffer the severe self-interference and need extra dedicated hardware and energy for the interference cancellations [10][11], which are big burdens for UAVs. To address such an issue, in our earlier works [12][13], a multi-UAV in-band full-duplex (MU-IBFD) aerial communication system with low hardware complexity is proposed for a multi-UAV video transmission system, in which both GS and UAVs are equipped with high-gain directional and beam steerable antennas so that the system can reuse radio resources and enable systematic full-duplex communication for multiple UAVs.

Experimental studies were also conducted. A testbed for UAV-to-car communication was built to measure the effects of UAV position and antenna orientation/location on performance [14]. A prototype of UAV controlled over LTE was demonstrated, and feasibility of the existing LTE as UAV communication infrastructure was evaluated in [15]. In [16], field measurements are conducted for UAVs connected to LTE networks, and the performance of massive UAV deployment is analyzed through simulations. In [17], a mmWave-based UAV communication system is built for raw 4K transmission. A comprehensive survey on prototypes and experiments of UAV communications can be found in [18].

In this work, preliminary experiments are designed and conducted to confirm the feasibility of the proposed multi-UAV IBFD system and show its key merit, i.e., the co-channel interference (CCI) cancellation among UAVs by directional antennas and UAV position control instead of energy-consuming dedicated cancellers on UAVs. Duo to the limitations of Radio Act Law and the regulations of university, a preliminary on-ground experiment is designed and conducted to reproduce the aerial propagation environment and confirm the system performance. The experiment is divided into outdoor channel power measurements and indoor communication performance measurements. This work in progress aims to provide preliminary results for the multi-UAV-based experiments in the air in the future.

The rest of this paper is organized as follows. Section II briefly recalls the system architecture. Section III describes the experimental system architecture and experimental procedure. The results are given and discussed in Section IV. Finally, Section V concludes this paper.

## II. SYSTEM ARCHITECTURE

For ease of analysis, the topology of interest in this work is the star topology for applications such as remote sensing and wireless relay. The proposed multi-UAV IBFD system is illustrated in Fig.2(a), and the architecture is shown in Fig.2(b). For simplicity, only two UAVs are illustrated in detail, because interference only occurs between two UAVs reusing same channels, and other pairs of UAVs follow the same mechanism. More details of MU-IBFD architecture can be found in [12] and [13]. In this paper, we only focus on the interference on the UAV side and the effect of this interference on system performance. Because the GS does not have the strict weight, size, and function constraints of UAVs, it is assumed to use conventional IBFD architecture, as shown on the right side of Fig.2(b).

In the system, the uplink and downlink of a UAV employ two different and sufficiently separated channels, so that complicated hardware for self-interference (SI) cancellation (SIC) in conventional IBFD architectures on a single UAV can be avoided, and the adjacent-channel interference (ACI) between Tx/Rx can also be effectively suppressed by analog filters. (In this paper, the uplink refers to a link from UAV to GS, and the downlink refers to the opposite.) To achieve high spectrum efficiency and realize multi-UAV IBFD, each channel is re-allocated to uplink of one UAV and downlink of another. For example, as shown in Fig.2(b), Channel#1 (blue) is re-used by the uplink of UAV#1 and the downlink of UAV#2. Similarly, Channel#2 (green) is re-used by downlink of UAV#1 and uplink of UAV#2. All other UAV pairs reuse the channels in the same mechanisms. Therefore, all channels in the system are simultaneously reused by uplinks and downlinks of multiple UAVs. The duplex and multiplex schemes of the proposed system are illustrated in Fig.2(c). Therefore, from the aspect of system, multi-UAV IBFD communication can be realized.

Because in this work we only focus on performance on the UAV side, i.e., downlink (GS to UAVs) performance, CCI between UAVs becomes the major issue affecting system performance. Two techniques are adopted jointly in the proposed system to address the problem. 1) *Directional antennas*: All UAVs are equipped with high-gain directional beam-steerable antennas and beamforming is performed, so that high isolation between Tx and Rx chains in propagation domain can be achieved passively without extra dedicated hardware and software cancellers on UAVs. In addition, the high directional gain also increases the received power. 2) *Controllability and mobility of multi-UAV systems*: To further increase isolation between the Tx/Rx chains in different UAVs reusing the same resources, the controllability, collaborativity, and mobility of multi-UAV systems are exploited, which are most distinctive features of multi-UAV system. Based on antenna directivities and positional relations of different UAVs, UAVs can use active position control to reduce CCI and ensure communication performance. Moreover, because multiple UAVs are typically operated in a collaborative manner, tasks can be dynamically assigned to the UAVs based on their positions and communication performance to guarantee the completion of global tasks.

## III. DESCRIPTION OF EXPERIMENT

Experiments are designed and conducted to reproduce the

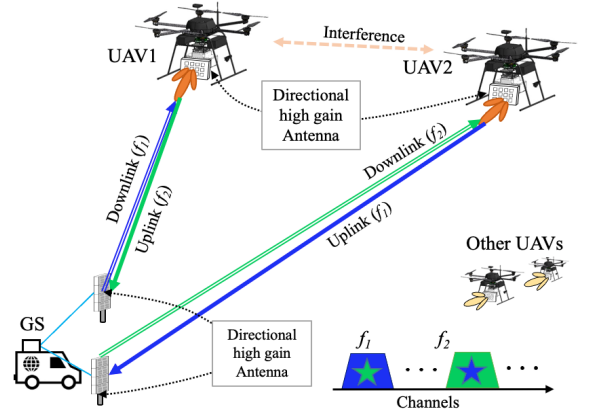
(a) Multi-UAV full-duplex communication system

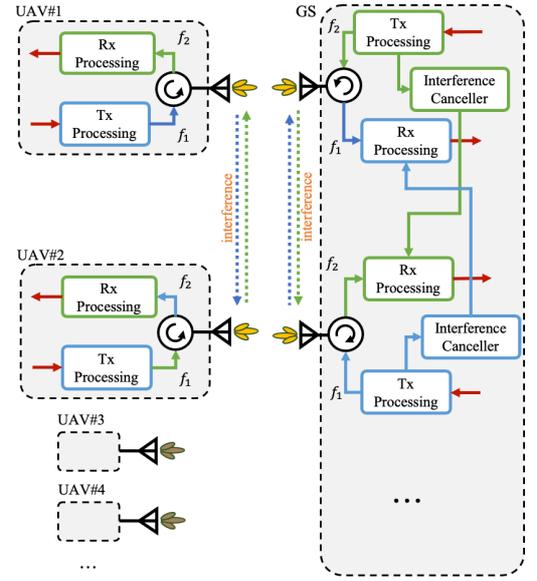
(b) System architecture

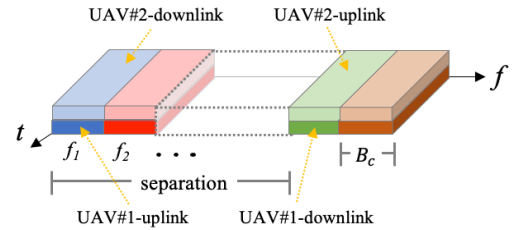
(b) Spectrum usage

Fig. 2 Architecture of the proposed multi-UAV communication system

application scene of the proposed system and explore its feasibility and effectiveness. To avoid the license application for flying refitted UAVs and sending experimental modulated signals to the air, the experiment is conducted on ground, and it is divided into outdoor channel power measurements and indoor communication performance measurements.

### A. Experiment Architecture

In the proposed multi-UAV communication system, on UAVs side, interference only occurs from uplinks to downlinks of two UAVs reusing the same channels, given that ACI is well canceled by filters. In addition, the interference to each other is symmetric and equivalent, if transmission powers of two UAVs are the same. Therefore, to explore the feasibility and effectiveness of the proposed system, an experimental prototype system is implemented in a minimum architecture as shown in Fig.3. In the experiment, Tx#2 at GS and Rx#2 at UAV#2 are implemented for downlink, and Tx#1 at UAV#1 is implemented to generate the interference signal from uplink to downlink. (Rx#1 at GS is not implemented because it is assumed to employ the conventional IBFD/SIC architecture and is out of the scope of this work.) All of them work on the same channel.

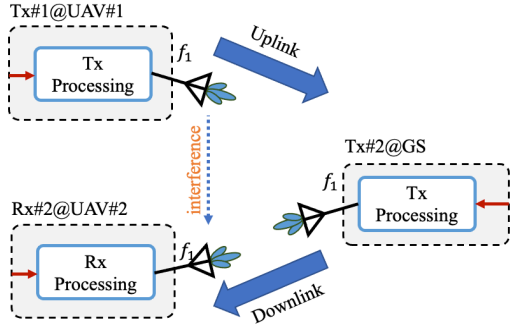

Fig. 3 Minimum experimental archtecture

In the prototype system, baseband processing is implemented in PCs, and radio frequency (RF) band processing is done in SDR devices. The experiment used high-gain directional horn antennas with a half-power beamwidth (HPBW) of 18° and a gain of 21dBi. For comparison, dipole antennas with a gain of 2.5dBi are also used.

### B. Interference Channel Power Measurement

Outdoor measurements are conducted to measure the power of channels, including the desired signal channel power from GS to UAV#2, i.e., $p_{G \to U2}$, and interference channel power from UAV#1 to UAV#2, i.e., $p_{U1 \to U2}$ in the proposed system. In the experiment, positions of GS (Tx#2) and UAV#2 (Rx#2) are fixed, and power of channels is measured while moving the position of UAV#1 (Tx#1).

The goal of this measurement is to reproduce the propagation environment for UAVs. One key difference between aerial and terrestrial environments for the proposed architecture is that the interference link is dominated by the line-of-sight (LOS) path between two UAVs in the aerial environment while it is dominated by the reflection path in the terrestrial environment. This difference also means that the proposed system is only available for aerial communication systems. To reproduce the aerial propagation environment, the measurements are conducted in a playground, as shown in Fig.4. The size of free space without surrounding buildings and trees is about 140m in front of Rx#2, and 60m on the left and right of Rx#2. According to our pre-test, reflection paths from Tx#1 to Rx#2 by the surroundings are negligible in the experiment due to the usage of directional antennas and the long distance to the surroundings. The playground's material is artificial grass, made of plastic and rubber. To eliminate ground reflections, the reception antenna of Rx#2 is placed tightly on the ground.

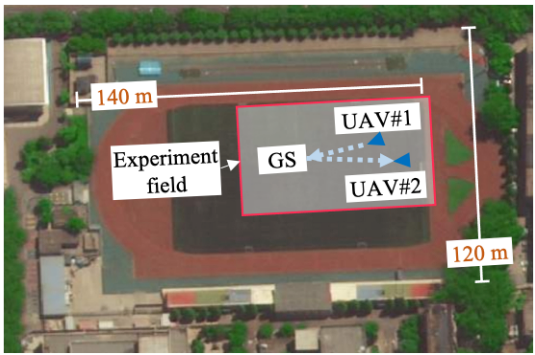

Fig. 4 Experiment envrionment of interference measurement

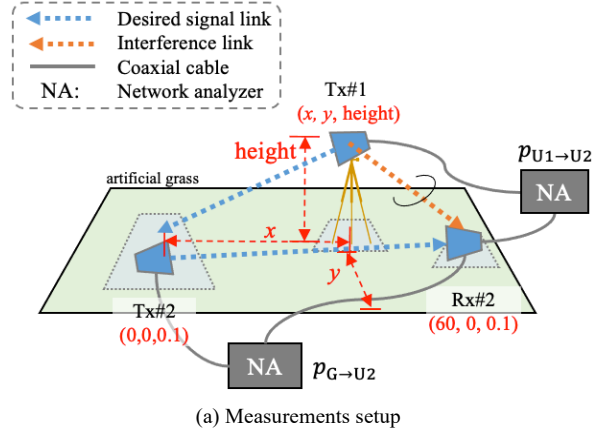

(a) Measurements setup

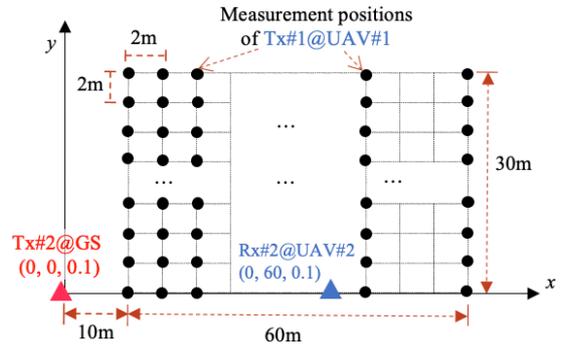

(b) Measurement points

Fig. 5 Experiment setup of interference measurement

The measurement is set up in the manner depicted in Fig.5(a). Tx#1's transmission antenna is put on a tripod for the convenience of adjusting the directivity and height. Beamforming is performed so that the main beams of Tx#2 and Rx#2's antennas face each other, and the main beam of Tx#1's antenna faces to the GS. However, it is noted that the beamforming is done manually in the experiment, aided by a laser pointer, so the effect of beam-misalignment could be included in results of interference channel power measurements because Tx#1's antenna direction needs to be re-adjusted in all measurement positions. The channel power of desired signal channel $p_{B \to U2}$, and interference channel power $p_{U1 \to U2}$ are measured by a network analyzer via coaxial cables.

The positions transmitters and receiver in the experiment is depicted in Fig.5(b). All coordinates are in meters. Tx#2 is put at a fixed position (0, 0, 0.1), and Rx#2 is put at fixed position (0, 60, 0.1), so that $p_{B \to U2}$ is measured only once. $p_{U1 \to U2}^{i}$ is remeasured every time when Tx#1 is in a new position, and $i$ is the index of positions. Tx#1 is moved every preamble insertion. Rx baseband consists of modules for synchronization, FFT, channel estimation, equalization, carrier de-mapping, IFFT, de-modulation, and FEC decoder. RF processing, including, e.g., digital-to-analog conversion (DAC)/analog-to-digital conversion (ADC), frequency up-conversion/down-conversion, is performed in USRPs. To simplify the complexity of implantation, three USRPs use the same external clock and pulse-per-second (PPS) signal. Experimental signals have bandwidth of 10MHz, modulation level of 16QAM, carrier frequency of 5.7GHz, and sampling rate of 15.36MHz. Two adjustable attenuators are used to generate path loss of desired signal channel and interference signal channel based on the channel power measurements.

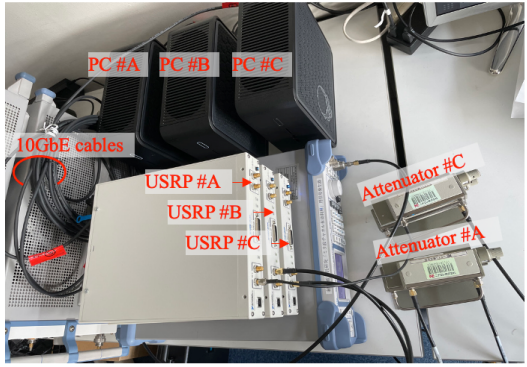

(a) Prototype hardware system

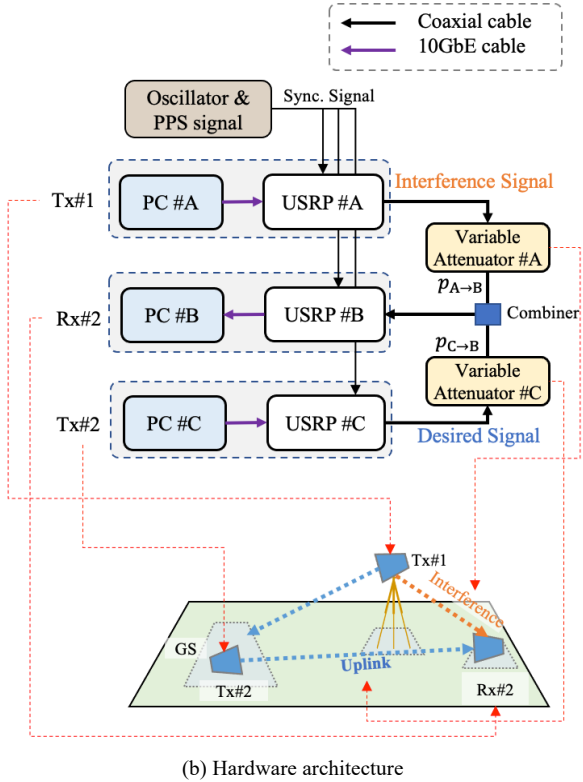

(b) Hardware architecture

Fig. 6 Experiment setup of channel capacity measurement

The architecture of the experimental hardware and connections between devices are shown in Fig.6(b). PC#A and USRP#A are used as Tx#1 to generate the interference signals. PC#C and USRP#C are used as Tx#2 to generate the desired signals. PC#B and USRP#B are used as Rx#2 for reception. After USRP#A's output (interference signals) passes through Attenuator#A, and USRP#C's output (desired signals) passes through Attenuator#C, they are fed to a signal combiner, and then the combined signal is fed to USRP#B as input. The attenuations, i.e., $p_{C \to B}$ and $p_{A \to B}^i$, are adjusted according to the channel power measurement results in all measurement positions to reproduce the path loss.

$$\begin{aligned} p_{A \to B}^i &= p_U + p_{U1 \to U2}^i \\ p_{C \to B} &= p_G + p_{G \to U2} \end{aligned} \quad (1)$$

where $p_U$ and $p_G$ are the assumed transmission power of UAVs (Tx#2) and GS (Tx#1).

In practice, the distance from GS to UAVs is typically very long and much longer than the distance between UAVs, so the reception power of desired signal in downlink is always very weak compared with transmission power of UAVs. To reproduce such a situation in experiment within the working range of hardware, $p_G$ is set to -45dBm and $p_U$ is set to 0dBm, when directional antennas are used. In addition, for comparison, when dipole antennas are used, $p_G$ is set to -8dBm and $p_U$ is set to 27.5dBm. It is noted that in real system $p_G$ is larger than $p_U$.

To evaluate the communication performance, signal to interference noise ratio (SINR) of received signals is calculated based on the root-mean-squared (RMS) error-vector-magnitude (EVM) at Rx#2 for all measurement positions. And then achievable channel capacity is calculated based on SINR. The measured capacity could be lower than capacity theoretically calculated based on the signal power, interference power, and theoretical noise power, due to the limitations of the performance of used prototype hardware such as amplifiers and ADC/DAC.

## IV. EXPERIMENT RESULTS

Experiment results are given and discussed in this section. In experiment interference power and communication performance in positions in positive y-axis, as shown in Fig.5(b), are measured, but a bilateral symmetry of the measurement results is also plotted on the negative y-axis for ease of understanding, because of the symmetric patterns of the antennas in experiment.

The results of power measurements of interference channel (Tx#1 to Rx#2) are given in Fig.7. The measured

2m from 10m to 70m in the x-axis, and every 2m from 0m to 30m in the y-axis, as shown in the Fig.5(b). In addition, measurements are also conducted when Tx#1 is at different heights, i.e., 0.1m and 1.8m.

### C. Communication Performance Measurement

By using the results from channel power measurements, the communication performance is measured by a prototype hardware system developed on SDR devices as shown in Fig.6(a). Three sets of universal software radio peripherals (USRPs) and host PCs are used as the two transmitters and one receiver in the minimal experiment architecture. Baseband processing is implemented on host PCs. It is noted that USPRs and PCs are only for performance confirmation in this preliminary on-ground experiment, and FPGA-based hardware is under development for the experiment employing UAVs. Tx baseband consists of modules for forward error correction (FEC) encoder, modulation, FFT, carrier mapping, pilot insertion, IFFT, CP insertion, and interference power, when Tx#1 is in different positions, is shown in Fig.7(a), (b) and (c) for cases in which a directional antenna at height of 0.1m, a directional antenna at height of 1.8m, and a dipole antenna at height of 0.1m are used, respectively. It is noted that Tx#2 is at (0, 0, 0.1) and Rx#2 is at (60, 0, 0.1). The lowest measured power in the figures has a floor around -95 dBm due to the sensitivity of the used spectrum analyzer. The fluctuations in results are caused by noise and the misalignment when performing manual beamforming in experiment.

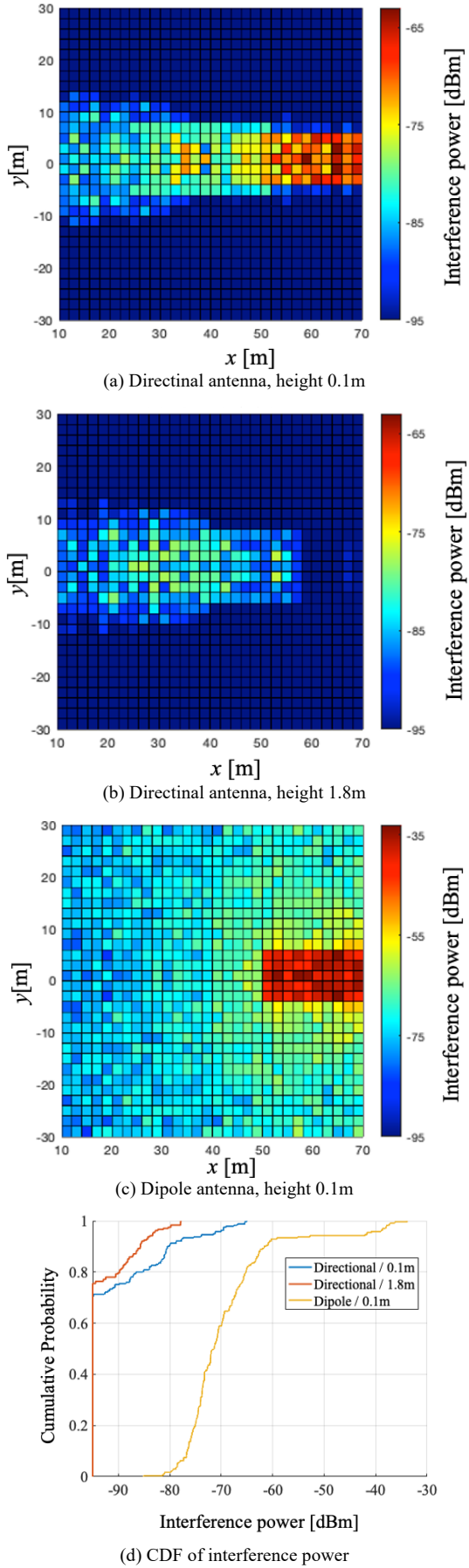

Fig. 7 Results of interference power

Fig.7(a) evaluates the scene that Tx#1, Tx#2 and Rx#2 with directional antennas are in the same plane. Obviously, it is the scene in which interference is the most severe. Fig.7(a) shows that large interference from Tx#1 to Rx#2 occurs when two UAVs are close to each other or in the main lobe direction of each other. Otherwise, much lower interference can be obtained. Interference channel power in around 71% of the experiment area is lower than -95dBm. Further, the scene that Tx#1 with directional antenna is at height of 1.8m is evaluated in Fig.7(b). Results show that due to the change in height, interference is greatly reduced, especially in positions near Rx#2. Interference channel power in around 76% of the experiment area is lower than -95dBm. Results in Fig.7(a) and (b) show that the distribution of interference power highly depends on the relative positions between Tx#1 and Rx#2. It proves that CCI can be eliminated by the combination of directional antenna and position control in the proposed architecture. For comparison, dipole antenna is also evaluated, and the results in Fig.7(c) clearly show that compared with results in the earlier two scenes, much larger interference occurs. Cumulative distribution functions (CDFs) of interference in above-mentioned three scenes are given in Fig.7(d), and significant improvement from the yellow curve to blue and red curves can be seen.

By manually adjusting the attenuations of the two adjustable attenuators, i.e., $p_{C \to B}$ and $p_{A \to B}$, based on the channel power measurements, the propagations of desired signal and interference signal are reproduced. The achievable channel capacity is calculated by the RMS-EVM measured in the prototype hardware system explained in Fig.6, when $p_G$ is set to -45dBm and $p_U$ is set to 0dBm for directional antenna, and when $p_G$ is set to -8dBm and $p_U$ is set to 27.5dBm for dipole antenna.

Achievable capacities are shown in Fig.8, in which results in Fig.(a) and (b) correspond to interference in Fig.7(a) and (b) with directional antenna at height of 0.1m and 1.8m, respectively. Results show that downlinks in the system can be established via the proposed architecture by directional antennas and position control. Given that uplinks in the system are established by using conventional IBFD and SIC architecture, the results indicate that multi-UAV IBFD communication can be successfully established. On the other hand, in case that dipole antenna at height of 0.1m is used, because the interference is too high, time synchronization in experiment even fails for all measured positions, which indicates that IBFD communication is impossible. For comparison, achievable channel capacity in case of time-division-duplex (TDD) OBFD system with 20% guard interval (GI) is calculated to be 11.6Mbps based on the measured signal power. Because of the orthogonal time slots, no interference occurs, so that archivable capacity does not

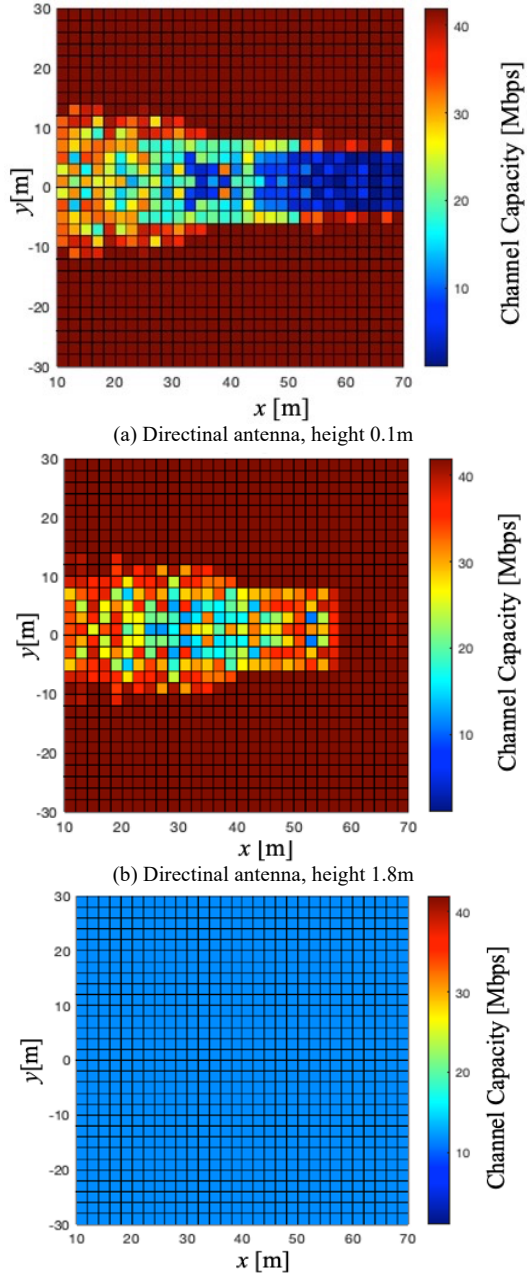

(a) Directinal antenna, height 0.1m

(b) Directinal antenna, height 1.8m

(c) Dipole antenna, height 0.1m (TDD)

Fig. 8 Results of achievable channel capacity

depend on positions, as shown in Fig.8(c), in which the scale of color bar keeps the same with Fig.(a) and (b) for ease of comparison. It shows that with the same hardware and experimental configurations, the proposed system can achieve around 4 times channel capacity in the experiment if position of Tx#1 is well controlled according to its relative positions with GS and Rx#2.

## V. Conclusion

To confirm the feasibility of the proposed multi-UAV full-duplex communication system equipped with directional antennas, an on-ground experiment is designed and conducted in this paper. The experiment results show that the power of CCI between two UAVs and the consequent achievable channel capacity highly depend on relative positional relationship of two UAVs. The results also show that much higher system performance can be achieved if the positions of two UAVs are well controlled. It demonstrates the effectiveness of CCI cancellation among UAVs using directional antennas and UAVs position control. More practical experiment by flying UAV in real environment is planned to be conduct soon.

ACKNOWLEDGMENT

This work is supported by Japanese Ministry of Internal Affairs and Communications (MIC) under the grant agreement 0155-0083.